\documentclass[preprint,proceedings]{rmaa}

\usepackage{paralist}

\usepackage{psfrag,color}

\SetYear{2004}
\SetConfTitle{Bodenheimer Fest}

\title{Destruction of Stellar Disks by Photoevaporation}

\author{
  D. Johnstone,\altaffilmark{1,2} 
  I. Matsuyama,\altaffilmark{3}
  I.G. McCarthy,\altaffilmark{2}
  and A.S. Font\altaffilmark{2} }

\altaffiltext{1}{National Research Council Canada, Herzberg Institute
of Astrophysics, Victoria, Canada}
\altaffiltext{2}{Department of Physics \& Astronomy, University of Victoria, Victoria, Canada}
\altaffiltext{3}{Department of Astronomy \& Astrophysics, University of Toronto, Toronto, Canada}

\shortauthor{Johnstone et al.}
\shorttitle{Photoevaporation of Stellar Disks}

\fulladdresses{
\item Andreea S. Font: Department of Physics \& Astronomy, University of Victoria, Victoria, 
                    BC, Canada, V8P 1A1.
\item Doug Johnstone: National Research Council Canada, Herzberg Institute of Astrophysics, 
                      5071 West Saanich Rd., Victoria, BC, Canada, V9E 2E7.
\item Isamu Matsuyama: Department of Astronomy and Astrophysics, University of Toronto,
                       Toronto, ON, Canada, M5S 3H8.
\item Ian G. McCarthy: Department of Physics \& Astronomy, University of Victoria, Victoria, 
                    BC, Canada, V8P 1A1.
}

\listofauthors{D. Johnstone, I. Matsuyama, I.G. McCarthy, \& A.S. Font}
\indexauthor{Johnstone, D.}
\indexauthor{Matsuyama, I.}
\indexauthor{McCarthy, I.G.}
\indexauthor{Font, A.S.}

\abstract{
Photoevaporation may provide an explanation for the short lifetimes of disks
around young stars. With the exception of neutral oxygen lines, the observed 
low-velocity forbidden line emission from T Tauri stars can be reproduced by 
photoevaporating models. The natural formation of a
gap in the disk at several AU due to photoevaporation and viscous
spreading provides a possible halting mechanism for migrating planets
and an explanation for the abundance of observed planets at these
radii.
}

\addkeyword{accretion}
\addkeyword{accretion disks}
\addkeyword{stars: formation}

\begin{document}
\maketitle

\section{Introduction}
\label{sec:intro}

Most low-mass stars form surrounded by circumstellar disks. 
Observations of infrared excess (e.g., Haisch, Lada, 
\& Lada 2001) suggest disk lifetimes 
$\tau_{\rm disk} \approx 6 \times 10^6\,$yr, requiring an efficient 
mechanism for disk dispersal.  

Hollenbach, Yorke, \& Johnstone (2000) considered a variety of disk dispersal
mechanisms and concluded that viscous accretion of the disk onto the central
star (e.g. Hartmann et al.\ 1998) together with photoevaporation of the disk
at moderate radii 
(Shu, Johnstone, \& Hollenbach 1993; Hollenbach et al.\ 1994; Johnstone, 
Hollenbach, \& Bally 1998) must act together efficiently to remove the entire 
disk. Numerical calculations by Clarke,
Gendrin, \& Sotomayor (2001) and Matsuyama, Johnstone, \& Hartmann (2003a)
have shown that these processes may remove the disk in $10^{5-7}\,$yr.

The photoevaporation model fits observational data well in the case of 
external heating via nearby massive
stars (Bally et al.\ 1998; Johnstone, Hollenbach, \& Bally 1998; St\"orzer \&
Hollenbach 1998).  With the exception of a few cases (e.g., MWC349A), 
evidence for disk photoevaporation due to the central
star (Shu et al.\ 1993; Hollenbach et al.\ 1994) is largely circumstantial.  
Observations of blue-shifted, low-velocity emission from forbidden lines
of oxygen, nitrogen, and sulfur in the spectra of many T Tauri stars 
(Hartigan, Edwards, \& Ghandour 1995) provide useful diagnostics.
Font et al.\ (2004) calculated the flow properties of the photoevaporative 
disk wind and found them to compare reasonably with the strengths and profiles 
of the nitrogen and sulfur lines. The oxygen lines, however, are 
underabundant in the model.

Along with disk dispersal, photoevaporation and viscous accretion 
produce structure in the disk such as gaps and rings. Matsuyama,
Johnstone, \& Murray (2003b) showed that 
the formation of gaps within the gaseous disk during the dispersal era 
places constraints on the migration of planetary orbits.

The following sections review the key results of Matsuyama et al.\ (2003a,b) 
and Font et al.\ (2004).

\section{Photoevaporating Disks}
\label{sec:evap}

The evolution of a stellar disk under the influence of viscous
accretion and photoevaporation from the central source or
external stars has been studied by Clarke et al.\ (2001) and 
Matsuyama et al.\ (2003a).  The key parameter in these studies is 
the source and strength of the ionizing radiation $\phi_*$. Clarke et al.\ 
assumed that the ionizing flux was due to super-solar activity on the surface
of the star and constant with time. Matsuyama et al., however, treated the 
photoionizing flux from the central source as arising from both 
the quiescent star and accretion shocks at the base of stellar 
magnetospheric columns. In the latter study, therefore, the ionizing flux 
was calculated self-consistently from the accretion mass-loss rate.

Alexander, Clarke, \& Pringle (2004) raised 
objections to the use of accretion energy to produce ultraviolet photons,
showing that such photons would be absorbed in the dense accretion flow and 
may not escape the shock front. Furthermore, both stellar and accretion shock
ultraviolet photons may be trapped by the strong protostellar jet
which launches from the inner disk. Shang et al.\ (2002) were unable
to ionize completely the jet using only photons from the central star due to
the high optical depth of the jet itself. It may be that the source
of ultraviolet radiation ionizing the disk is the disk-jet 
boundary, in which case the photon flux still scales as the accretion energy.
Other possibilities include enhanced stellar surface activity 
(but only in stars with weak or absent jets) or  nearby massive stars.

The disk cannot be entirely removed using only viscous accretion and 
photoionization from the disk-star accretion shock because the 
photoionizing flux decreases too quickly with time 
(Matsuyama et al.\ 2003a). When 6-13.6 eV far-ultraviolet (FUV) photons
from relatively nearby massive stars can be included, however, the disk 
is removed in \( 10^{6-7}\,\)yr. In addition, when ionizing photons
from close massive stars can be included the disk is removed in 
\( 10^{5-6}\,\)yr. 

\begin{figure}[!t]
  \includegraphics[width=\columnwidth]{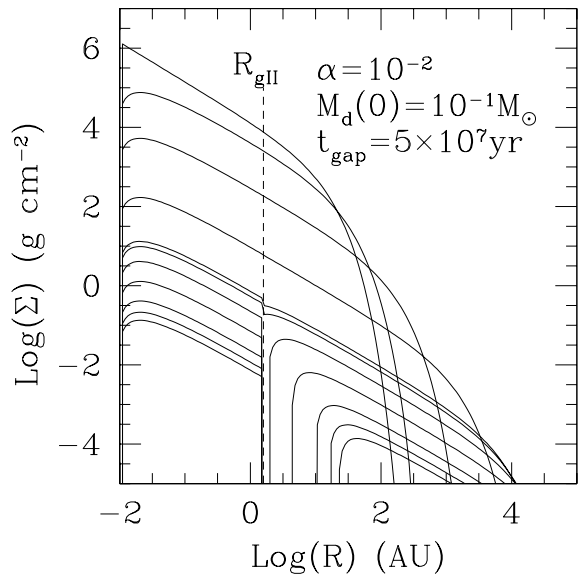}
  \caption{Snapshots of the disk surface density for a representative model
with both viscous evolution and photoevaporation from the central star. $R_{gII}$
refers the gravitational radius for $10^4\,$K gas. See Matsuyama et al.\ 2003a for 
details.} 
\label{fig:gap}
\end{figure}

An intriguing consequence of photoevaporation by the central star is the
formation of a gap in the disk at late stages of the disk evolution
(Figure \ref{fig:gap}). 
The gap forms near $r_g \sim G\,M_*/a^2$, where the sound speed $a$ of the 
$10^4\,$K photoevaporated
disk gas matches the escape velocity from the central star. 
Interior to this radius, material from the disk is bound to 
the system and forms a hot disk corona. Exterior to this radius, the ionizing
radiation is less intense due to divergence. The gap forms when
the surface density at $r_g$ drops to $\Sigma(r_g) \sim 2\,$g\,cm$^{-2}$,
for typical ionization and viscous parameters.

\begin{figure}[!t]
\includegraphics[width=\columnwidth]{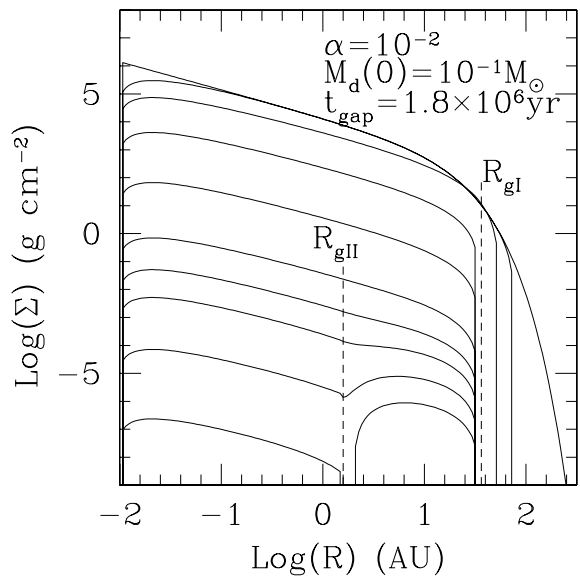}
\caption{Snapshots of the disk surface density with viscous evolution, 
internal photoevaporation, and evaporation due to a nearby massive star.  
$R_{gI}$ refers the gravitational radius for $10^3\,$K gas. See Matsuyama 
et al.\ 2003a for details.}
\label{fig:ext}
\end{figure}

When viscous accretion and photoevaporation by both 
the central star and nearby massive stars are considered simultaneously, 
the disk shrinks and is truncated at an outer radius, set by
the temperature in the FUV photoevaporated disk gas
($T \sim 10^3\,$K) or the temperature in the ionizing photon 
photoevaporated flow ($T \sim 10^4\,$K). 
At typical distances (\( d\gg 0.03\, \)pc) from 
massive stars the disk is evaporated by both external FUV photons 
and internal ionizing photons. This produces both an outer 
limit for the disk size and
an inner disk gap during the last stages of disk evolution
(Figure \ref{fig:ext}).

\section{Hydrodynamic Models}
\label{sec:hydro}

Low-velocity (e.g. $\sim 10\,$km\,s$^{-1}$) 
blue-shifted forbidden lines are observed
in the spectra of T Tauri stars (Hartigan et al.\ 1995).  The 
similarity between the blue-shift of the low-velocity component and the sound
speed in $10^4\,$K gas has led to speculation that this feature may be an 
observational signature of photoionized disk winds.

\begin{figure}[!t]
  \includegraphics[bb=40 10 450 420 ,clip,width=\columnwidth]{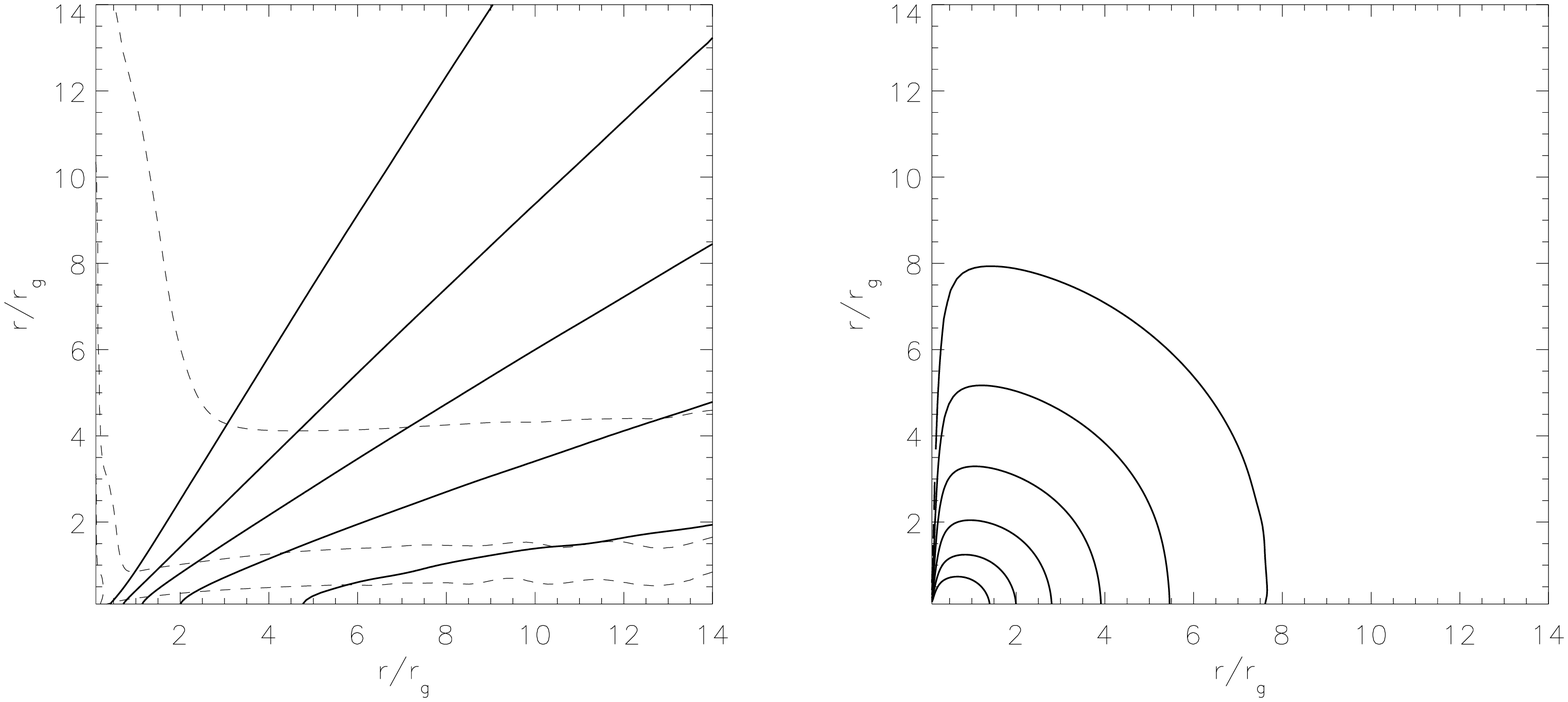}
  \caption{Steady-state flow results for the photoevaporative disk wind model.
  The solid lines are streamlines and indicate how the ionized gas travels. The
  dashed lines indicate where $v_{\rm tot}/a$ equals 1, 2, and 3. See
  Font et al.\ (2004) for details.}
\label{fig:flow}
\end{figure}

Using the analytic model of Shu, Johnstone, \& Hollenbach (1993) and Hollenbach
et al.\ (1994) as a basis, Font et al.\ (2004) examined the characteristics of 
photoevaporative outflows using hydrodynamic simulations.  
Figure\,\ref{fig:flow} shows the streamlines and isotachs of the resultant disk 
wind.  Near the disk surface, the radial density gradient forces the streamlines
to bend outward significantly and the flow accelerates. At larger distances 
from the disk, however, the flow becomes radial and approaches the classical 
Parker wind solution (Parker 1963).  

General results from the simulations were found to agree well with 
the analytic predictions, although some small differences were present. Most 
importantly, the flow of material from the disk surface develops at a somewhat
smaller radius than in the analytic approximations ($r_g = G\,M_*/3\,a^2$) and 
the flow-velocity from the disk surface is only one-third the sound speed.  

\begin{figure}[!t]
\includegraphics[width=\columnwidth]{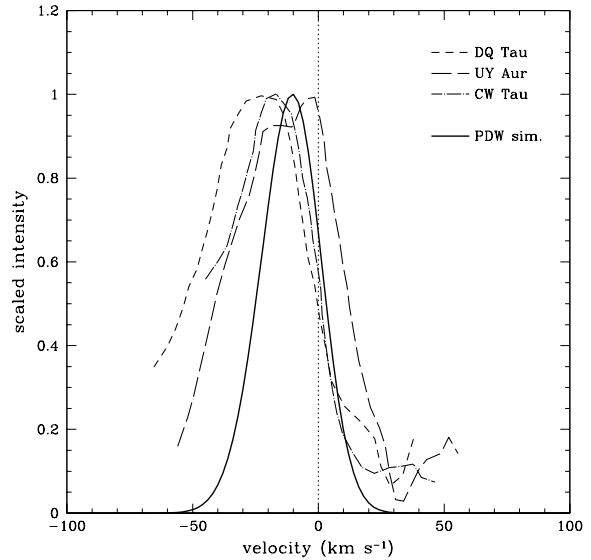}
\caption{Comparison of [SII]$\lambda$6731 line profiles. The 
photoevaporative disk wind (PDW) model profile is calculated assuming
$\phi_* = 10^{41}\,$s$^{-1}$, $M_* = 1\,M_\odot$ and a face-on disk.
See Font et al.\ (2004) for details.} 
\label{fig:profile}
\end{figure}

Assuming the gas to be almost entirely 
ionized and adding contributions from each cell in the hydrodynamic simulation,
Font et al.\ (2004) computed the widths, velocities, and luminosities of
sulfur and nitrogen forbidden lines. 
The simulated line shapes were found to be in relatively 
good agreement with observations (Figure \ref{fig:profile}).

Font et al.\ (2004) also demonstrated that the model line strengths are in
agreement with observed low-velocity forbidden line
emission of ionized species from T Tauri stars.  This is in contrast with
magnetic wind models (Garcia et al.\ 2001a, 2001b), which systematically 
under-predict these line luminosities.  The photoevaporative model, however, 
cannot account for the luminosities of 
neutral oxygen lines in T Tauri stars since almost all oxygen
in the model is ionized. 

\section{Migration of Planetary Orbits}
\label{sec:planets}

The recent discovery of Jupiter-mass planets orbiting at 3-5 AU from their 
stars complements earlier detections of massive planets with very small orbits. 
The short period orbits suggest that migration has occurred, 
possibly due to tidal interactions between the planets and 
the gas disk. The newly discovered long period 
planets (and  our own solar system) show that 
migration is either absent or rapidly halted in at least some systems. 

\begin{figure}[!t]
\includegraphics[width=\columnwidth]{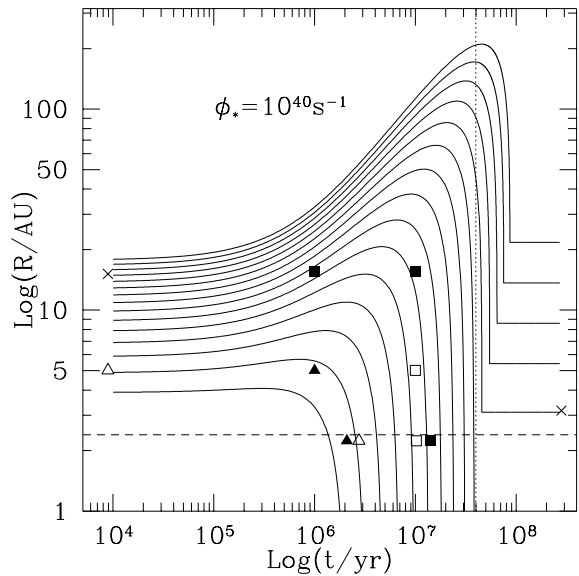}
\caption{Planet streamlines in a viscously evolving and photoevaporating disk.
See Matsuyama et al.\ (2003b) for details.
} 
\label{fig:migrate}
\end{figure}

Matsuyama et al.\ (2003b) proposed a novel mechanism for halting 
disk coupled planet migration at several AU in a gas disk.  As mentioned 
in \S \ref{sec:evap},
photoevaporation of the disk may
produce a gap in the disk at $r_g$ (e.g. a few AU). Such a gap would prevent 
outer planets from migrating inward toward the star, resulting in an 
excess of systems with planets at or just outside $r_g$. 
Figure \ref{fig:migrate} plots the streamlines of planets moving 
in tandem with their viscously evolving disk. Planets
which form in the inner disk migrate with the viscous disk toward the
central star. In contrast, planets from the outer disk
are halted by the gap near $r_g$, fixing their orbits at several AU.
Almost all planets which tidally lock themselves to the disk early will
first migrate outward in the disk before viscously accreting toward the
star.

\section{Conclusions}
\label{sec:conc}
Photoevaporation of disks around young stars may be responsible for
short observed disk lifetimes, especially when powered by FUV  or ionizing
radiation from nearby massive stars. Together, photoevaporation and 
viscous accretion naturally lead to the formation of gaps and ring 
structures within disks, without the need to invoke unseen planets.

Hydrodynamic models of the photoevaporative disk wind reveal that
the launching point for the flow, $r_g$, has been overestimated in the
analytic models by a factor of $\sim\,3$ and that the launch velocity
has been also somewhat overestimated.  While most observed low-velocity
forbidden line profiles are reasonably matched by the model, the predicted 
neutral oxygen forbidden line flux is too low since almost all oxygen in
the model is ionized.  Interestingly, a hot $10^4\,$K wind
which is not photoionized would contain mostly neutral oxygen and the line
strengths would produce an excellent fit to the observations. 

The existence of a gap in the disk at $r_g$ provides a halting mechanism
for migrating planets. Given the predicted location of $r_g$ at several 
AU from the central star, such gaps might 
explain the recent abundance of observed planets
with these radii.

\acknowledgements

DJ would like to thank Peter Bodenheimer for many interesting discussions
on star and planet formation. D.\ Hollenbach and F.\ Shu have influenced
a great deal of this work. The authors would also like to thank L.\ Hartmann, 
N.\ Murray, and D.\ Ballantyne for their scientific input during these 
studies.  Thanks to J.\ Di Francesco for a critical reading of this manuscript.

\end{document}